\documentclass[aps,prb,superscriptaddress,showpacs,twocolumn]{revtex4}

\usepackage{amssymb}
\usepackage{graphicx}

\begin{document}
\title{Similarity and contrasts between thermodynamic properties at
   the critical point of liquid alkali metals and of electron-hole droplets}
\author{F.E. Leys}
\affiliation{Antwerp University (RUCA), Groenenborgerlaan 171, B-2020
   Antwerpen, Belgium}
\author{N.H. March}
\affiliation{Antwerp University (RUCA), Groenenborgerlaan 171, B-2020
   Antwerpen, Belgium}
\affiliation{Oxford University, Oxford, England}
\author{G.G.N. Angilella}
\affiliation{Dipartimento di Fisica e Astronomia, Universit\`a di
   Catania, and Istituto Nazionale per la Fisica della Materia, UdR
   Catania, 64, Via S. Sofia, I-95123 Catania, Italy}
\author{M.L. Zhang}
\affiliation{Antwerp University (RUCA), Groenenborgerlaan 171, B-2020
   Antwerpen, Belgium}

\begin{abstract}
The recent experimental study by means of time-resolved luminescence
measurements of an electron-hole liquid (EHL) in diamond by Shimano et al.
[Phys.Rev.Lett. {\bf 88}, 057404 (2002)] prompts us to compare and contrast
critical temperature $T_{c}$ and critical density $n_{c}$ relations in
liquid alkali metals with those in electron-hole liquids. The conclusion
drawn is that these systems have similarities with regard to critical
properties. In both cases the critical temperature is related to the cube
root of the critical density. The existence of this relation is traced to
Coulomb interactions and to systematic trends in the dielectric constant of
the electron-hole systems.  Finally a brief comparison between the alkalis
and EHLs of the critical values for the compressibility ratio $Z_{c}$ is
also given.\\
\end{abstract}

\date{}

\maketitle

Two of us\cite{me} have been concerned in earlier work with the behavior of
the thermodynamic quantities $T_{c}$, $n_{c}$ and $p_{c}$, at the critical
points of the fluid alkali metals. Relationships were shown to exist between
critical temperature $T_{c}$ and critical number density $n_{c}$ for the
five fluid alkalis, the form anticipated in the earlier investigation of
Chapman and March \cite{Chap}, namely 
\begin{equation}
T_{c}n_{c}^{-\frac{1}{3}}=\mathrm{constant}  \label{alk}
\end{equation}
being confirmed in \cite{me}.

The present study has been prompted by the very recent experiment of Shimano
et al. \cite{Shim}, who report on the formation of an EHL \cite{gene} with a
high critical temperature $T_{c}=165K$ in diamond, by means of
time-resolved luminescence measurements under an intense femtosecond
photoexcitation above the band gap. Then, by time-resolved spectral shape
analysis, a very high carrier density $n_{0}=1.0\times 10^{20}$cm$^{-3}$ at 
$T=0$ is revealed, together with the high value of the critical
   temperature $T_{c}$ already recorded above. 

In addition to the important findings for the EHL in diamond, Shimano et al. 
\cite{Shim} comment in their introduction on the similarities between the
physical properties of the EHL and liquefied metals. It is their comment
which has motivated the present study, in which we compare and contrast EHL
properties with those of the fluid alkali metals. The Eq.~(\ref{alk}) above
provides a natural starting point, but unfortunately while $T_{c}$ is known
for at least 5 EHLs, the same is not true for the critical density $n_{c}$.

\begin{table}[b]
\begin{tabular}{cccc}
\hline
 & $n_0$ (cm$^{-3}$) & $T_c$ (K) & $\varepsilon$ \\
\hline
Ge & $2.5 \cdot 10^{17}$ & 6.7 & 16.0 \\
Si & $3.3 \cdot 10^{18}$ & 24.5 & 12 \\
GaP & $6 \cdot 10^{18}$ & 40 & 9.1 \\
3C--SiC & $7.8 \cdot 10^{18}$ & 41 & 9.72 \\
Diamond & $1.0 \cdot 10^{20}$ & 165 & 5.7\\
\hline
\end{tabular}
\caption{Values for the EHL number density $n_{0}$ at $T=0,$ the critical
temperature $T_{c}$ and the dielectric constant $\varepsilon $ in five
indirect-gap semiconductors (Extracted from Table 1 of Ref. 3).}
\end{table}

However, prompted by the form (\ref{alk}), we have taken data from
   Ref.~\cite{Shim}, and have collected $T_{c}$ and the zero
   temperature densities $n_{0}$ for $5$ EHLs in Table~1 of the
   present paper. From these values, it can 
be readily shown that whereas $T_{c}$ increases by a factor $\sim 25$ in
going from Ge to diamond, the zero temperature quantity $n_{0}$ increases by
a factor $4\times 10^{2}$. Due to the form (\ref{alk}), which we emphasize
however contains the critical density $n_{c}$ in the fluid alkalis, we have
studied first the question as to whether $T_{c}$ correlates with $n_{0}$ in
the $5$ EHLs referred to in Table~1. After some numerical investigation, we
have constructed Fig.~1, which demonstrates beyond reasonable doubt
   that $T_{c}$ and $n_{0}$ indeed correlate strongly. There is power
   law behavior 
between $T_{c}$ and $n_{0},$ although the $\frac{1}{3}$ power in Eq.~(\ref{alk}) must be modified (empirically to $\frac{1}{2}$: see caption to
Fig.~1 and the further comments below). However, while Fig.~1 displays
similarities of shape with Eq.~(\ref{alk}) for the fluid alkalis, this
equation contains the critical density $n_{c}$ whereas for lack of
experimental information, Fig.~1 correlates $T_{c}$ with the zero temperature
density $n_{0}$. 

\begin{figure}[t]
\centering
\includegraphics[height=\columnwidth,angle=-90]{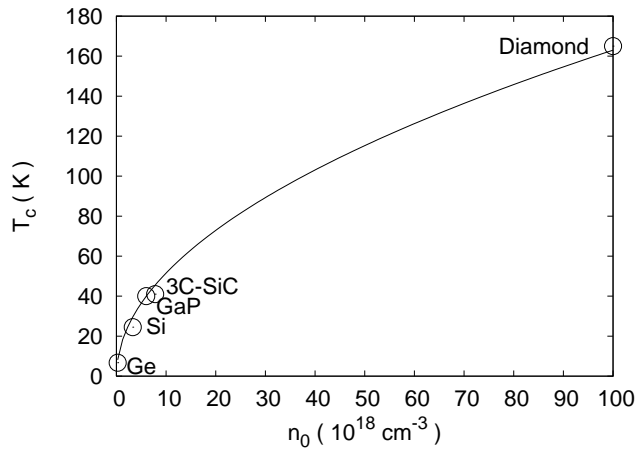}
\caption{Critical temperature $T_{c}$ of EHL in the five semiconductors
referred to in Table~1 versus the (zero temperature) number density $n_{0}$.
The curve drawn is given by $T_{c}n_{0}^{-\frac{1}{2}}=16.3\times
10^{-9}cm^{\frac{3}{2}}$~K.}
\end{figure}

In the absence of experimental information on $n_{c}$, we have sought a
theoretical basis for bringing Eq.~(\ref{alk}) and Fig.~1 into closer contact.
We are considerably aided here by the work of Kalt et al. \cite{Kalt}. These
authors observed picosecond electron-hole droplet formation in the indirect
gap material $Al_{x}Ga_{1-x}As$. In the course of their experimental work,
they refer to a 'scaling law' 
\begin{equation}
\frac{n_{c}}{n_{0}}\simeq 0.3  \label{scal}
\end{equation}
with a reference to Forchel et al. \cite{For}. Obviously, if we adopt this
so-called 'scaling law', then Eq.~(\ref{alk}) and the result from Fig.~1 come
into intimate context, and the conclusion is that there is a different
exponent, say generally denoted by $\varkappa$, for the relationship 
\begin{equation}
T_{c}n_{c}^{-\varkappa }=\text{constant}  \label{gen}
\end{equation}
with $\varkappa =\frac{1}{3}$ for the $5$ fluid alkalis and $\sim
   \frac{1}{2}$ from Fig.~1 and Eq.~(\ref{scal}) for the
   EHLs. Already, Chapman and March 
\cite{Chap} had noted the totally different critical behavior for liquid
alkali metals compared with condensed rare gases, for the latter case the
exponent $\varkappa$ being $-2$.

As this brief report was nearing completion, we became aware of two further
contributions. In the first of these, written more than $20$ years ago by
Reinecke and Ying \cite{Rein}, the authors had anticipated a relation
   $T_{c}n_{c}^{-\frac{1}{2}}=\mathrm{constant}$. This is the more
   remarkable because 
it can be seen from the present Fig.~1 that the 'diamond' point of Shimano et
al. \cite{Shim} is crucial to fitting the parabolic form, and allowing the
above constant to be made wholly quantitative. The writers who anticipated
the parabolic relation, however, went on to express doubts as to whether
such a relation had any fundamental basis. This leads us to the second
contribution referred to above: that of Likal'ter \cite{Lik}.

Likal'ter studies what he emphasizes is the limiting situation of the EHL in
which the hole mass $\ m_{h} \gg m_{e},$ the electron mass. He then gives
individual formulae for the critical constants $T_{c}$ and $n_{c}$ 
discussed above, and also for the critical pressure $p_{c}$. Our interest in
Likal'ter's formulae is to expose relations between these critical constants
which can then be compared and contrasted with known results for the fluid
alkali metals.

\begin{figure}
\centering
\includegraphics[height=\columnwidth,angle=-90]{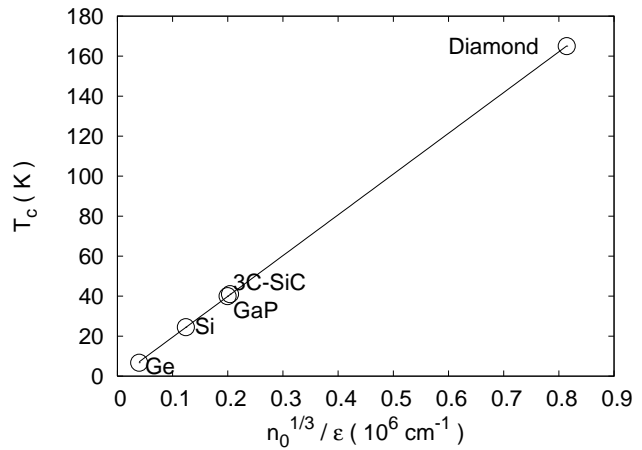}
\caption{Critical temperature $T_{c}$ of the EHL of five semiconductors
referred to in Table~1 versus $\frac{n_{0}^{\frac{1}{3}}}{\varepsilon}$,
where $n_{0}$ is the number density of the EHL at $T=0$ and $\varepsilon $
the dielectric constant.}
\end{figure}

The first $T_{c}$--$n_{c}$ relation which follows from Likal'ter's model is 
\begin{equation}
T_{c}=\left[ \frac{4}{7}\right] ^{2}\left[ \frac{4\pi }{3}\right] ^{1/3}
\frac{e^{2}\gamma }{2}\frac{n_{c}^{1/3}}{\varepsilon }  \label{lika}
\end{equation}
where $\gamma$ is a renormalized Madelung constant, given by Likal'ter
   as $\approx 0.55.$ This formula differs qualitatively from the
   fluid metals 
finding of Chapman and March\cite{Chap} by the dielectric constant
   $\varepsilon$ of the semiconductors appearing in the
   denominator. The 
limiting prediction of Eq.~(\ref{lika}) deduced from Likal'ter's model will
now be brought into close contact with the plot in Fig.~1. Thus in Fig.~2 we
have used the data from Table~1, including the experimental dielectric
constant values recorded in the final column, to plot $T_{c}$
   vs. $\frac{n_{0}^{\frac{1}{3}}}{\varepsilon }$. There is clearly a
   linear relation as 
predicted by Likal'ter's model. Accepting the scaling relation (\ref{scal}),
the parabolic fit of the data in Fig.~1 can be reconciled with the linearity
of Fig.~2 provided $\varepsilon$ and $n_{0}$ are related, of course
approximately, by a $\frac{1}{6}$ power law. However, at present we have no
fundamental justification for a relation $\varepsilon
   n_{0}^{\frac{1}{6}}=\mathrm{constant}$, because, at least in
   principle, such a formula could contain the 
binding energy of excitons, or the effective electron mass.

Returning to the similarities and differences from the fluid metals, we have
also employed Likal'ter's individual formulae for $T_{c}$, $n_{c}$ and
   $p_{c}$ to calculate the so called compressibility ratio $Z_{c}$
   defined as  
\begin{equation}
Z_{c}=\frac{p_{c}}{n_{c}k_{B}T_{c}}  \label{zc}
\end{equation}
which was studied in detail for the five fluid alkalis in Ref.~\cite{me}. The
limiting formulae (for $m_{h} \gg m_{e}$) then lead to the result 
\begin{equation}
Z_{c}=\frac{7}{24}  \label{zval}
\end{equation}
This is near to the value for the heavier alkalis, $0.217$ for Rb and $0.203$
for Cs, but there is a wide variation through the alkali fluids, Li having a
value of $0.064$. The value in Eq.~(\ref{zval}) is not so much smaller than
the prediction from the van der Waals equation of state prediction,
   namely $3/8$ \cite{Vdw}.

The conclusion is firstly that there are some interesting similarities
between the EHLs and the fluid alkali metals with regard to critical
properties. In both cases $T_{c}$ is related to $n_{c}^{1/3}$, which
reflects the fingerprints of Coulomb interactions. For the EHL however, in
the limiting case studied by Likal'ter \cite{Lik}\thinspace where
   $m_{h} \gg m_{e},$ this quantity $n_{c}^{1/3}$ is divided by the dielectric
constant $\varepsilon$ of the semiconductors, high for Si and considerably
reduced for diamond. Secondly, the limiting formulae of Likal'ter for
   $T_{c}$, $n_{c}$ and $p_{c}$ are shown to yield the constant critical
compressiblility ratio $Z_{c}$ in Eq.~(\ref{zval}), in contrast to the wide
spread of values of $Z_{c}$ in the fluid alkalis.

N.H.M. wishes to acknowledge some financial support from the Francqui
Foundation. Thanks are also due to the Executive Director, Professor L.
Eyckmanns, for his help and encouragement.
G.G.N.A thanks the University of Antwerp (RUCA) for much hospitality.
Finally, the authors wish to acknowledge the constructive comments made by
two referees, which have enabled us to improve the presentation of this
Report.

\end{document}